# Balancing freedoms, rights and responsibilities during COVID in US: a study of anti- and pro-restriction discourse

Gillian Bolsover
University of Leeds
g.bolsover@leeds.ac.uk

**Abstract**

*Countries across the world have instituted unprecedented restrictions on freedom of movement, privacy and individual rights to control the spread of COVID-19. These measures tend to have been derived from communally orientated East Asian cultures. The way that culturally relevant concepts of rights and freedoms underpin COVID restrictions in democratic and individually orientated countries remains unknown. This data memo addresses this issue through analysis of pro- and anti-restriction discourse on social media in the US. It finds that anti-social and economic restriction discourse more frequently articulates rights and freedoms, based on ideas of inviolable rights to freedom of movement or freedom of economic activity or a cost-benefit analysis that places economic activity over public health. Pro-social and economic restriction discourse less frequently mentions rights and freedoms, instead supporting restrictions as following state and medical advice and out of deference and respect to medical professionals. Discourse is highly polarised and divisive and articulated largely through established political identity positions. It is suggested that more attention is paid to discussions of balancing rights and freedoms in COVID control restrictions. To convince opposers of restrictions, supporters of restrictions should base arguments around communal rights and positive freedoms. It is also important to critically evaluate whether and how these perspectives need to be adapted to be appropriate and resonant in democratic and individualistic countries.*

**Restricting rights and freedoms to fight COVID**

COVID-19, a pandemic with recorded cases in almost every country, was first identified in the Chinese city of Wuhan in December 2019. In late January, with human-to-human transmission confirmed and facing the Chinese New Year travel period, China instituted an unprecedented lockdown of Wuhan and the wider Hubei Province (BBC, 2020a). Hubei's 60 million residents were subject to strict lockdown measures, with individuals forbidden to leave their apartments without permission and community officials delivering goods and medication. A variety of enforcement methods were employed across China with some localities financially compensating those who reported others who broke lockdown rules and many localities using phone apps to track the movement of citizens both within and between jurisdictions (Feng & Chen, 2020).

As the virus has spread across the globe, many other affected countries have instituted similar kinds of restrictions on privacy and freedom of movement to control the spread of the disease. In Spain, citizens were not allowed to leave their homes, except for work or buying essential supplies and medicines (BBC, 2020b). For six weeks Spanish children under the age of 14 were not allowed out of the house (Hedgecoe, 2020). In France, individuals had to carry a completed paper or app-based declaration that stated one's reasons (from an approved list) for being in public and were subject to fines if they could not produce this declaration (Broom, 2020). Various degrees of restriction have been instituted in almost every affected country.

These measures represent unprecedented trade-offs between different kinds of rights and freedoms, including economic v. public health consequences, freedom of movement v. freedom from disease, protecting vulnerable members of society v. restricting children's access to education, socialisation and play. Notably these measures were first employed in China, where an authoritarian state and collectively orientated culture, would tend to lead to different ways of balancing rights and freedoms compared to democratic and individually orientated societies.

Different states have also made different choices in terms of the extent to which balancing risks has been made the responsibility of individual citizens. Sweden initially chose not to impose restrictions, with prime minister, Stefan Löfven arguing that the government could not legislate and ban everything but rather that individuals should take responsibility for their own health and the health of the community, using their own common sense to regulate their behaviours (Orange, 2020). Similarly, when the UK moved from the slogan "Stay at Home" to "Stay Alert" some argued that this was an attempt to shift responsibility, blame and the burden of managing these risks and restrictions from the state to individuals (Jones, 2020).

These debates have been particularly prominent in the US, the country most affected by the virus with almost 4 million cases and 145,000 deaths as of 24 July 2020 (CDC, 2020). Coronavirus responses have been piecemeal in the US, showcasing a lack of clarity about whether the federal or state government is responsible and with President Donald Trump ordering states that have imposed restrictions to reopen their economies (Selin, 2020).

With an individualistic culture, a high proportion of libertarian supporters and a constitution strongly protecting individual freedoms, the US has seen several major protests against state-level economic and social restrictions to prevent the spread of COVID-19. A particularly notable protest occurred in Michigan on 30 April, when hundreds of armed protestors entered the capitol building during a debate over Democratic governor Gretchen Whitmer's request to extend the emergency powers that underpinned the state's stay-at-home measures (Beckett, 2020). Photos of heavily armed protestors in combat gear inside the state capitol kicked off a discussion in the US about how rights and freedoms were being balanced in local and national responses to the COVID pandemic.

**Balancing rights and freedoms in political thought and discussion**

A commonly used distinction in political thought is the difference between negative and positive liberty, or freedom from and freedom to (Berlin, 1969). In this dichotomy, the negative concept of liberty is understood as the freedom to act without interference (such as freedom of assembly or freedom of speech). In contrast, the positive concept of liberty is understood as the freedom from a source of control or interference (such as freedom from discrimination or freedom from violence)[1]. Heated political debates often arise over how negative and positive freedoms are balanced, such as between the right to bear arms in the US and the right to be free of violence or between the right to freedom of speech and the right to be free from hateful attacks.

Public discussions around coronavirus responses have also often discussed the balancing of economic and public health (or welfare) rights, with lockdowns to control the disease having a negative impact on economic activity (Portes, 2020). A third pole of these debates are individual human rights to privacy and freedom of movement, which are also negatively impacted by COVID-controlling measures. Commentators have spoken out against hastily constructed movement tracking and content tracing initiatives that many fear will represent a permanent encroachment on individual rights and spill over into non-COVID uses (EFF, 2020).

This type of right's balancing has long been a concern for scholars of Chinese politics, with Chinese political systems seen as prioritising welfare rights (such as freedom from hunger and poverty) over individual human rights (such as freedom of speech and freedom of movement) (Perry, 2008; Zhou, 2010). China, and other East Asian cultures, are also seen as collective, more willing to restrict individual liberty and freedom to protect community health (Hofstede et al., 2010). These countries also have a much stronger emphasis on protection of the elderly through the Confucian concept of filial piety.

China's authoritarian state also has a political and practical system that facilitates COVID restrictions, with an existing infrastructure for tracking that other countries are only now trying to build to respond to COVID. There are existing controls on freedom of movement in China through the household registration (Hukou) system as well as the world's most advanced system of technological surveillance and control.

While there has been little debate about COVID restrictions in China, the implementation of restrictions has triggered intense debates in democratic contexts, particularly in the US with its individual and rights-orientated political system and fragmented federal system. These debates have played out across traditional media, through numerous anti-lockdown protests and on social media, which individuals increasingly rely on to get their political information, express their opinions about politics and organise collective political action. Given how new and unprecedented these restrictions have been, public opinion and reaction to these initiatives (as well as deeper opinions on how rights and freedoms should be balanced in a global pandemic) are as yet unknown.

**Research Approach**

Responding to this provocation, this data memo will present research into how individuals are discussing the balancing of rights and freedoms in the context of COVID within the US. In order to do this, the research will focus on discourse on the social media platform Twitter. Although Twitter is not the largest social media platform in the US, it is the one most focused on news and current affairs. It is also the most open major social media platform and, therefore, an appropriate venue for studying a public discourse ecosystem rather than communications within delineated groups.

There has been some survey research about people's opinions in this area, such as work from the Pew Research Centre that found 60% of US respondents thought that if the government tracked people's locations through their cellphone, it wouldn't make much of a difference in limiting the spread of COVID-19 (Auxier, 2020) and Associated Press polling that found 80% of US respondents supported measures that advise people to stay at home and limit social interactions to 10 of fewer people (Rambaran, 2020). However, while these surveys provide a quantified assessment of opinions in the population, they cannot encapsulate the landscape of debates nor what articulation of a balance between different rights and freedoms underlies the opinions people express in these surveys. An analysis of social media can help answer these questions.

Given the unprecedented and fast-moving nature of COVID policies and practices, it is extremely important to understand how members of the public are reacting to these policies and how their reactions are underpinned by different cost-benefit analyses of how rights and freedoms should be balanced in this circumstance. In order to address this question and to contribute to ongoing discussions about this issue, this data memo will present analysis of discussion on Twitter data in the US during the week of the 27 April through 5 May. This week-long period was selected as it contained the 30 April Michigan state capitol protest.

Many research projects intending to monitor social media discourse on Twitter collect data from within a group of pre-selected hashtags and keywords. However, this strategy risks missing emergent or unselected topics. It is thus severely limited in its ability to speak to the body of online discourse, particularly during fast-moving events, and is subject to significant researcher bias based on the selection of hashtags and keywords to follow. To avoid this limitation, this project collected a sample of data from all trending topics within the US during the study period.

Using custom Python scripts to interface with the Twitter API, the project collected the most recent 100 tweets associated with each of the top 50 trending topics in the 64 locations for which Twitter collates trends (including one for the entire country) every 15 minutes during the target period: 27 April through 5 May 2020. This data collection captured 3,421 unique trending topics across the week-long period.

A content analysis was then undertaken of the 200 most popular of these trends to determine their topical content, based on the majority of tweets made within that trend during the time-period. The topical coding scheme used an established coding frame for social media posts developed by the researcher and used in previous studies in the US and China (Bolsover, 2017, 2018) and in previous research about COVID in the US (Bolsover & Tokitsu Tizon, 2020).

Within the top 200 trends, 38 (19%) concerned COVID, of which 28 (14%) were political commentary on COVID policies and practices. Intercoder reliability tests were performed on a set of trends collected using the same method during a different period: 23 – 26 April (Bolsover & Tokitsu Tizon, 2020). Percentage agreement for being about COVID was 93.8% with a Kappa of 77.8%. Percentage agreement for being political commentary on COVID policies and practices was 96.7%[2]. These reliabilities are well within established appropriate ranges for this type of research (Lombard et al., 2002).

Out of these top 200 trends, only one trend concerned anti-lockdown protests anywhere in the US. For this reason, in addition to assessing the most popular trending topics, we briefly considered each of the 3,421 trends to assess whether it concerned anti- or pro-lockdown discourse or the protests in Michigan specifically. We then randomly selected 100 tweets from within each of four groups of trending topics:
- the 27 trends about COVID policies and practices from within the top 200 trends (excluding the one about anti-lockdown protests);
- the five trends from within the whole dataset that were specifically designed to express anti-lockdown messages;
- the three trends from within the whole dataset that were specifically designed to express pro-lockdown messages;
- the three trends from within the whole dataset that specifically concerned the Michigan state capitol protests.

**Table 1: Tweet dataset for analysis**

|  | No. of trends | Total no. tweets in trends | No. of tweets analysed |
|---|---|---|---|
| COVID policies and practices | 27 | 467,709 | 100 |
| Anti-lockdown trends | 5 | 20,413 | 100 |
| Pro-lockdown trends | 3 | 32,382 | 100 |
| Michigan protest trends | 3 | 11,691 | 100 |

This spread of 400 tweets across four different groups of trends was designed to provide a picture of the general nature of commentary on COVID policies and practices on US Twitter, specific pro- and anti-lockdown discourse and discourse about the Michigan protests. We then analyse these posts to understand how individuals expressed their opinions of how rights and freedoms should be balanced, and indeed, which rights and freedoms deserve discussion.

**Overview of pro- and anti- lockdown attitudes**

Firstly, it should be noted that, as shown in Table 1, there was more activity in openly pro-lockdown trends than in openly anti-lockdown trends. However, this activity is not necessarily indications of actual levels of support as the analysis of tweets from within these groups shows significant pro-lockdown activity in anti-lockdown trends and vice versa.

It is also important to note that not all posts from within these trends concerned the coronavirus. From the general COVID policies and practices, only 40 (out of 100) concerned COVID. From within anti-lockdown trends this was 98; pro-lockdown 72 and Michigan protest 66. The fact that irrelevant content was much more common in the more generalised, more popular trending topics has interesting implications for future research using the search function of the Twitter API to collect posts on a certain topic[3].

Trends specifically about the Michigan protests evidenced the largest number of posts articulating ideas of rights and freedoms (59 out of 100 examined posts). Posts in anti-lockdown hashtags also contained numerous posts articulating ideas of rights and freedoms (41 out of 100). However, ideas of rights and freedoms were less commonly articulated in posts in pro-lockdown hashtags (8) or in posts in trends concerning commentary on COVID policies and practices (18). This suggests that pro-lockdown messages may not be being articulated in terms of rights and freedoms (such as the right to a safe workplace or freedom from dangerous disease).

Although the pro- and anti-lockdown keyword sets clearly articulated a position, in actuality support for restrictions to control the spread of COVID was mixed across all the four groups examined. Within the 100 posts in trends commenting on COVID policies and practices, support and opposition of restrictions was fairly evenly balanced; seven posts supported restrictions on social activity, nine restrictions on economic activity and four restrictions on political activity. In contrast, eight opposed restrictions on social activity, 13 opposed restrictions on economic activity and seven opposed restrictions on political activity.

Within anti-lockdown hashtags, more posts opposed restrictions, but with a significant proportion that supported restrictions; 31 opposed restrictions on social activity, 32 opposed restrictions on economic activity and ten opposed restrictions on political activity. In contrast, 14 posts supported restrictions on social activity, 15 restrictions on economic activity and one restrictions on political activity. We see that the focus of these discussions was on economic and social restrictions, rather than political restrictions, despite that fact that discussions about the right to protest and about potential impacts on the upcoming 2020 election have been prominent in state and traditional media discourse.

Within pro-lockdown hashtags, there was less balance between supportive and opposing posts, with the bulk of posts supporting restrictions and a few posts opposing restrictions; 30 posts supported restrictions on social activity, 27 restrictions on economic activity and two restrictions on political activity. In contrast, five posts supported restrictions on social activity, five restrictions on economic activity and one restrictions on political activity.

Within trends concerning the Michigan protests specifically, there was more pro-restriction than anti-restriction content, with a sizeable minority opposing restrictions; 17 posts supported restrictions on social activity, 18 restrictions on economic activity and 24 restrictions on political activity. In contrast, seven posts supported restrictions on social activity, nine restrictions on economic activity and eight restrictions on political activity. The focus on political activity here, compared to little discussion in other trends, was largely centred on the right to protest and forms of legitimate protest. However, although these data can provide a broad overview of attitudes towards these restrictions on social media it is necessary to look at the content of these posts to understand how balances of rights and freedoms were being articulated.

**The meaning of freedom**

Across the dataset of 400 posts, 61 were opposed to either economic or social restrictions or both. Of those posts that articulated this opposition in terms of rights and freedoms, there were two clear groups. The first group expressed these views based on a conception of fundamental human rights, centred around freedom of movement. The second groups expressed these views in terms of economic rights.

Posts in the first group expressed anti-lockdown opinions according to an idea of inviolable freedoms and rights, understood as a freedom of movement. One post likened Michigan governor Gretchen Whitmer to Hitler. This concept was closely linked to ideas of patriotism. Pro- and anti-lockdown voices labelled the Michigan capitol protestors through a similar lens, as either Michigan patriots or Michigan terrorists, which will be discussed in more detail later. This approach to the virus centred around specific definitions of America or what it means to be American, such as:

*So if you don't do exactly as we say in #Chicago and #StayHome you will 'go to jail, period' and we will treat you like a criminal. However, criminals are being released from jail because of possible #COVID infections This is not the America I remember - what has happened?*

*#OpenCalifornia #opencalifornianow it's time people of the great nation of America to open your doors and not let a silly virus stop you!*

*I'm tired of pastors getting arrested for having church services I'm sick of hard-working Americans getting ticketed for going to a park I'm tired of good patriots getting destroyed while actual criminals walk freely. It's time for Justice. JUSTICE & TRUTH will UNITE America*

Within these posts we see ideas that the responses to control the coronavirus are un-American, un-patriotic and have targeted American values (such as freedom of religion and freedom of movement). These posts clearly see individual freedom of movement as the ultimate expression of freedom and rights, indeed this was what was meant by rights and freedoms in this context.

*The left is literally trying to Villainize Michigan Protesters because they are standing up for their rights, and are tired of their Governor nullifying the Bill of Rights.. Anyone who's against these people standing up to Tyranny is an Enemy to the Republic! #MichiganProtest*

In post below, #halfwhitmer is a reference to Gretchen Whitmer the Governor of Michigan whose attempts to extend the emergency bill that allowed stay at home orders the Michigan capitol protests targeted.

*#halfwhitmer just doesn't know when to stop!! Enough is enough!! #openmichigannow She wants to be considered for VP & doesn't care what she'll destroy to get it!! Fellow Michiganders we need to fight for our rights!! #MichiganProtest #MichiganPatriots #FreeMichigan #VoteRed2020*

This group of posts also exhibited ideas that the Coronavirus was being exaggerated for political purposes (by liberals or the Democratic Party) and linked this in a rather unspecified way to a Chinese political manoeuvring.

*China has a plan. Get the Liberals to #ExtendTheLockdown*

*The Democrats milked this Pandemic like a cow and got away with loading the Bills meant to help We the People with all the PORK they couldn't get on a stand alone Bill. If any American ever believed they truly care about them, they certainly shouldn't now.*

Several posts exhibited polarised political positions, such as linking the lockdown to debates about the right to choose to terminate a pregnancy. This suggests that debate about appropriate coronavirus responses has been polarised into existing political camps in the US, with opinions becoming an identity position rather than a debate informed by evidence and reason.

*Does anyone else find it ironic that we are all on lockdown, and businesses are closed, just to try and save lives, yet abortion clinics are still open? #openohionow*

This identity positioning was also evidenced in pro-lockdown tweets.

*Republicans claim to be pro-life while continuing to offer elder Americans up for sacrifice. You cannot make this up.*

Across the posts there was a high level of divisiveness. Out of the 400 posts, 166 appeared to be attempting to prevent others from speaking or undermine the value of their words; 95 saw distinct opposing groups in society (other than political parties) and 91 directed hate against one of these groups. Thus, 96% of posts that saw distinct opposing groups in society (91 out of 95) directed hate at the opposing group.

This divisiveness was similar across pro- and anti-lockdown content; 39% of posts that were anti-restriction saw distinct opposing groups and all but one of these posts directed hate at that group (or groups) (23 out of 24). For pro-restriction posts, 40% saw distinct groups and all but one of these posts directed hate at that group (or groups) (29 out of 30). Hate direction from one side was much more common within the hashtags that represent the ideas of that side; 65% of hate direction in anti-restriction posts occurred in the anti-restriction hashtags and 52% of hate direction in pro-restriction posts occurred in pro-restriction hashtags. This suggests that posters are more likely to express hateful content when they feel they are speaking to or within groups and topics that support those views. Genuine discussion may, therefore, be better found in more neutral hashtags. However, as previously mentioned these hashtags and keywords contained much less COVID relevant content, which would be detrimental to the ability to engage in cross-opinion discussion in these spaces.

Posters also evidenced high levels of partisanship. Out of the 61 posts that were anti-social or economic restrictions, 48% were made by accounts that were right-wing partisan individuals, who post content that is overwhelmingly dedicated to supporting right-wing parties and policy positions and with the identity presentation of the account obviously aligned as a partisan supporter. Out of the 75 posts that were pro-social or economic restrictions, 68% were made by left-wing partisan individual accounts. Interestingly, 20% of anti-social or economic restrictions posts were made by left-wing partisan individual accounts. This again suggests that a large proportion of those who post and engage in debate on these issues on Twitter are fighting for their chosen partisan position rather than engaging in genuine discussion and debate.

In sum, in this first group of posts, anti-lockdown sentiment is expressed in terms of a negative conception of rights and in particular an equation of ideas of freedom and rights specifically to a freedom of movement. These posts understand restrictions to control the coronavirus as un-patriotic and fundamentally un-American. This fits into established politically polarised positions, rather than ongoing debates based on evidence and reason. These posts included accusations of political manoeuvring and conspiracy by liberals, the democratic party and China. They were also linked to pre-existing identity position debates such as Christian identity, abortion and (as will be discussed later) gun ownership. Posts, in general, evidenced high levels of divisiveness and hate, particularly within hashtags that specifically expressed a polarised position on the debate.

**Economic calculations**

A second group of anti-lockdown posters expressed their opinions through a focus on economic and welfare rights, specifically the right to work, to shelter and to freedom from hunger and poverty. Although this group also evidenced significant political polarisation and political positioning, more posts in this group evidenced an approach based on balance and reason, coming to the conclusion that the economic harms of restrictions outweighed the public health harms of disease.

*I just wanna be able to go to work man. Bein poor sucks. #ReOpenOregon*

*Don't bankrupt hospitals, dry up our food supply, and destroy the lives and the health of Americans. #ReOpenOregon now!*

It is important, however, to place these posts within a political context. Although there is an unavoidable economic impact of restrictions, who is most greatly impacted is a matter of political policy. Some posts in this group pointed out that large and online businesses were thriving in these conditions, sometimes alleging conspiracy and corruption in coronavirus restriction efforts.

*Had to stop at the local Walmart today. It was PACKED. Just goes to show you how unethical this lockdown is. I can go to a packed Walmart to buy art supplies but I cant go to the tiny local art store to buy art supplies. The disproportionate hit to small businesses is criminal.*

Another post shared a headline, image and link to the blog of conspiracy theorist David Icke: *Jeff Bezos's Net Worth Has Increased by $24 Billion During the alleged 'Covid-19 Pandemic.'* The image showed Amazon founder and CEO Jeff Bezos smiling broadly with the words:

*Lockdown increases Bezos wealth by tens of billions, who would have guessed that Amazon would benefit from hundreds of millions of other businesses closing down worldwide – many never to reopen? The Bezos-owned Washington Posts says "Lockdowns must continue."*

A pro-restriction voice expressed similar sentiment in another way.

*It's wild to me that people think lockdowns are corporate/government conspiracies. The conspiracy you should be worried about is one where everyone is told that it's safe to resume daily life as usual, when it's not, because fucking Walmart is worried about their bottom line.*

Although there is some conspiracy content to these posts, there is an important truth here that some businesses have thrived and profited during COVID restrictions. Many of these have been large businesses that as individual entities are tautologously more 'essential'; tend to have larger resources to enable them to implement distancing and hygiene restrictions; and tend to have larger buildings that can adapt to social distancing restrictions in a way that small operations are much less able to. The owners and decisionmakers in these businesses are also at a greater distance to those who undertake practical labour and thus (although this is clearly not true of all businesses) may be more willing to subject employees to risk by keeping the business open. Whether those most able and those who have continued to grow and profit during COVID pay more to recovery and aid is very much an open question.

However, it is important to note a contradiction in the extension of the arguments made in this group of posts, if discussions around COVID responses continue to adhere to polarised identity positions. The following post demonstrates this perspective. Sharing a link to the far right[4] Gateway Pundit with the headline '*JUST IN: Northern California County Defies Governor Newsom, Reopens Churches, Hair Salons, Schools and Restaurants*', the post states:

*Wonderful. The people have had it with the socialist tyrants. Revolution is brewing. But I say FIRST we go to Supreme Court and test their violations of US Constitution. No Governor has right to order u or me to shut our business down.*

Within polarised positions, the argument against social and economic restrictions to fight COVID aligns with the perspective that state aid to assist those most affected and most unable to weather these restrictions is unconstitutional and unsupportable socialism. These positions are underpinned by a strongly individualistic interpretation of rights and freedoms and the role of governance.

**Individual vs. collective rights**

Within the anti-economic restriction posts, several posts argued that only those who advocate restrictions should be bound by them, expressing a wholly individualised approach.

*No one who wants to prevent others from working should be allowed to buy anything. Period. There's nothing magic about a grocery store or the instacart gig worker that brings groceries to you that's any safer than a hundred other kinds of work. #ReopenOregon #ReopenAmerica*

*If you're saying we should #ExtendTheLockdown will you sign this pledge? Wuhan Flu Lockdown Supporter. To demonstrate my commitment to suppressing the freedoms of my fellow Americans and robbing them of the dignity of earning a day's wage, I make the following binding unilateral contract: (1) I pledge to quit my current job so that someone in a "non-essential" profession can take it over. (2) If I can still keep a roof over my head, I pledge to abandon my home and to donate it to a family that is losing their home. (3) If I can still keep my car, I pledge to give that care to someone who has lost theirs. (4) If I can still put food on my table, I pledge to give that food to someone who is starving.*

This perspective is, perhaps, best expressed in a clip from a video interview of the Michigan capitol protestors shares by one poster, who states:

*Michigan Patriots are Awesome🇺🇸🔥 #Trump2020 #MichiganProtest*
In the video shared by the poster, a woman argues: *I know there's people that are compromised immune-wise and people that are older, weaker and they can't take it. They are free to stay in their homes. That's what freedom is. This isn't freedom.*

Here we see a primacy of individual rather than collective rights. This view has a limited and negative view of social responsibility (i.e. a responsibility not to interfere with others' freedoms) rather than a positive view of social responsibility as a set of actions to protect and aid vulnerable members of the community. This is a major reason why it is important to evaluate and discuss COVID responses as a balance of rights and freedoms.

Lockdown restrictions have been successfully implemented not just in China but in other collectively orientated countries in East Asia that share a Confucian history that places great stock in the protection of elder members of the community[5]. Mask wearing has also been commonplace for many years in several East Asian countries, in particular in dense urban areas, where people wear masks to avoid infecting others when they are feeling unwell[6].

It is notable that only one of the 75 posts that expressed support of economic or social restrictions or both phrased this support around ideas of social responsibility and protection of vulnerable members of the community.

*#DoNotOpenCalifornia OR ~~~ We need to protect our loved ones.* An image attached to the post reads*: I see a lot of people being like "I would survive the coronavirus. I'm taking my chances". The way I see it, yeah. I'd survive it. But I might carry it to someone who wouldn't. And that, folks, is the problem.*

Instead of positive freedoms and positive ideas of social responsibility, posts that supported economic or social restrictions tended either to take a politically polarised view that assumed the validity of restrictions as a foregone conclusion (and often took those who opposed them as idiots) or to phrase this support as a deference or act of respect to doctors and the medical profession.

**Lack of rights discourse in restriction supporting posts**

The first main group of posts that supported economic and social restrictions did so without evidence or reference to a discussion of a balance of rights and freedoms. There were a large number of these posts (22 out of the 75 restriction-supporting posts were selected as potential examples). Typical posts of this type were:

*ya'll better stop with this #opencalifornianow bullshit, we're not ready to open for your fucking hiking and beach selfies, stay the fuck at home and wash your nasty hands*

*If you attended the #ReOpenOregon protest today and you align with the proud boys, QANON, boogaloo's or other flat Earth thinking mentality. Fuck off and unfollow me. PS tell me your leaving so I can block you on the way out*

This again demonstrates the extent to which this issue (at least in the US) has become a politically polarised position issue position, which is largely not amenable to debate. It is important to note that this kind of issue polarisation is incompatible with the ideas of community responsibility and communal rights that underpin the logic of restrictions to control the spread of COVID.

Another main group of posts within restriction supporting posts largely phrased their support around respect for or deference to healthcare workers.

*My good friend is an emergency room nurse in Central California. She just saw that #opencalifornianow was trending and begged me to ask my followers to get #DoNotOpenCalifornia to spread. She said it's horrific there.*

*stop with #opencalifornianow. just because \*\*only\*\* 1,800 people DIED so far? we've spoken to so many doctors, hospitals, the WHO, and they are SLAMMED, working beyond overtime. do not make this worse for your fucking pinkberry and hiking selfies, stay in your fuckin house.*

*#DoNotOpenCalifornia Show some fucking respect for the Drs, nurses, and other health care personnel who are destroying themselves physically and emotionally to keep COVID-19 patients alive!! Don't flood them with more sick and dying patients. SO SELFISH!!*

While it is impossible to deny that respect and support is due to healthcare workers at this time, the use of this as the main reason for support of social and economic restrictions risks sidelining some of the important rights and freedoms-based arguments in support of these restrictions. One important reason for these restrictions was to prevent health services from being overwhelmed; however, community social responsibilities and positive freedoms (such as the right to be free from disease and the right not to be exposed to undue risk in the workplace) were also important considerations. These arguments are perhaps better counters to the individual freedom-based arguments of freedom of movement focused supporters. The focus in justification for these restrictions on preventing health services from being overwhelmed has also led to counterarguments that posit exaggeration and hoax in the threat and level of COVID deaths:

*Better question, why have flu deaths gone drastically down? It can't be because hospitals are being dishonest to get more Covid-19 money, can it? #OpenOhioNow*

*California has no Covid-19 Outbreak. Out of a Population of 38 Million only 1,800 deaths, most are from nursing homes! #OpenCaliforniaNow*

In sum, the 75 posts from the dataset that supported economic and/or social restrictions to control the spread of COVID in the US were less likely to express their support as an argument about balancing rights and freedoms, or even as pertaining to rights and freedoms. In comparison, the 61 posts that opposed economic and/or social restrictions more frequently expressed their views either according to ideas of an individual's inviolable right to freedom of movement, individual's inviolable right to economic activity or a cost-benefit calculation that favoured economic activity over protecting public health.

In total, 77% of the posts that opposed economic and/or social restrictions articulated ideas of rights and freedoms compared to 43% of posts that supported economic and/or social restrictions. This disparity is potentially because those who oppose these restrictions see themselves as fighting the state and, as such, express themselves according to the principles that underpin governance. In contrast, those who support the restrictions see themselves as following the advice or orders of states and professionals and, as such, do not make the arguments or counterarguments for the support of these restrictions independently.

It might be more productive for debate, however, if supporters expressed their views using language of rights and freedoms (with media and opinions leaders facilitating this discussion by discussing the issues through this lens). This is important not just around the issue of economic restrictions where discussions of balance have tended to centre but even more importantly around social restrictions and discussions of individual vs. community rights and the rights and freedoms an individual gains as a member of a community.

**Political rights and restrictions: Michigan case**

This data memo has thus far discussed individual expressions of support and opposition to economic and social restrictions on Twitter and the way in which these were expressed in terms of different rights and freedoms. We also looked in the dataset at expressions of support and opposition for political restrictions. Many of these discussions centered around the armed Michigan capitol protests on 30 April, with 31 posts expressing support for restrictions on political activity and 26 opposition.

These posts demonstrated high levels of political polarisation. The protests were viewed through existing political lenses, with supporters referring to the protestors as Michigan patriots and opposers as Michigan terrorists. As such, debate about these positions is largely precluded. Much more so than social or economic restrictions, ideas of political restrictions to fight COVID or the impact of social restrictions on political activity are an important issue that deserves much greater attention. This is because it is only through political activity that someone could register opposition to economic or social restrictions. If the need for restrictions restricts protest against those restrictions, the rational, limits and paths to effective alternatives need to be clearly communicated. Ideas of patriots or terrorists present foregone conclusions to that debate.

Another aspect of this polarisation was the fact that essentially all opposition to the Michigan protests was seen through the lens of two existingly polarised issues in American politics: gun ownership and race. (Messages of support for the protestors have largely already been covered in previous sections.)

A number of posts suggested that the kind of protest engaged in was illegitimate political activity (and that the individuals were not protestors but terrorists).

*Protesters should carry signs, not semiautomatics. #MichiganProtest*

*#MichiganProtest Heavily armed people storming a government building is an attack, not a protest. It is at the very least an act of intimidation, if not terrorism.*

*Dear #MichiganTerrorists You have a 1A right to peaceably assemble You have NO right to storm a State building with assault weapons, threaten peace officers or Governor Whitmer. You have NO right to infect others with COVID19. Arrest these terorists!*

Several posts suggested that this protest activity, which they saw as illegitimate, was instigated by the Republican Party and/or President Trump. One post showed a screenshot of a tweet by Trump saying *LIBERATE MICHIGAN* with the caption *Anyone know where the #MichiganTerrorists could've gotten the idea to storm the Capitol building?*

Another wrote:

*#Michiganders time to kick out the GOP. They are fomenting civil disorder and hurting Michiganders. Stand by #GovWhitmer and dump #GOP and #DumpTrump2020*

Supporters countered that it was, in fact, democrats, liberals and left-wing groups who use illegitimate protest techniques.

*This is what Democrats do to protest. Break windows, start fires and act like reckless criminal vigilantes. Republicans in Michigan today protested peacefully by occupying a building & chanting "this is the peoples house" Big difference. #MichiganProtest*

*The dumbest talking point is that the #MichiganProtest people are allowed to do this because they're white conservatives. BLM staged riots in multiple cities, set buildings on fire, looted. Hardly any arrests. Antifa shuts down streets, hits people with pipes. Hardly any arrests.*

This second post refers to the very common theme in Michigan protest opposing posts that the rights that were afforded to the protestors to protest during COVID, to protest in the capitol building and to carry weapons were racially (and to a lesser extent politically) orientated. They argued that these rights would not have been afforded if the protestors were not white. Many posts made these kinds of argument, of which some examples were:

*As a man who identifies as both Black and Muslim, I can guarantee that if I walked up into a State building armed to the teeth with a gripe, my Black Muslim ass would be shot on site, no questions asked. #MichiganProtest*

*Ask a group of African Americans to attend the legislature with high powered firearms and storm the house floor.....suddenly the 2nd amendment will be...flexible... #MichiganTerrorists*

*Armed white men are allowed to basically hold a Governor hostage and they're treated like they have rights while an unarmed Black child is brutally attacked for having a tobacco product. This is America. #MichiganProtest #MichiganTerrorists #COVID__19*

These posts discussed rights and freedoms but did so in terms of underlying and long-term issues of racism and police brutality in the US, without any mention of COVID. This touches on a greater discussion (that is outside the bounds of this short data memo) about how rights, freedoms and responsibilities are unequally applied and that structures and processes exist that mean that the rights of some groups are more infringed on that the rights of others. This data memo has been primarily concerned with articulations of balances of rights and freedoms that would underlie equally applied policies. However, this discussion of rights and freedoms at a more theoretical level would likely lead to greater discussion at the practical level of their application and the structural level of their interaction.

**Limitations and further directions**

This data memo has presented an initial analysis of levels of support and opposition to social, economic and political restrictions to control the spread and severity of COVID and how these arguments have been articulated in relation to rights and freedoms. An understanding of how individuals articulate these positions is particularly important in assessing, evaluating and crafting policy in this unprecedented situation.

This research found a rough balance between supporting and opposing opinions, although in each case posts supporting restrictions outnumbered posts opposing. In total, 68 posts supported social restrictions, 69 economic restrictions and 31 political restrictions. For opposition, these were 51, 59 and 26, respectively.

Posts that opposed social, economic and/or political restrictions were more likely to articulate ideas of rights and freedoms: 75% of posts compared to 50% of those that supported. Supporters articulated this either as with equating ideas of freedoms and rights to an inviolable right to freedom of movement; inviolable right to freedom of economic activity; or as an economic balancing act that favoured a continuation of economic activity over protecting public health. In contrast, when opposers mentioned rights it was overwhelmingly in the context of an uneven application of rights and freedoms in relation to the Michigan capitol protests. Posts supporting restrictions on economic or social rights tended either to justify this based on an assumptive correctness of the restrictions (with those who oppose being stupid or uninformed) or through ideas of deference and protection of medical professionals, upon whose shoulders treatment and care of COVID patients falls.

An analysis of these posts finds that, like the previously published analysis of health misinformation (Bolsover & Tokitsu Tizon, 2020), COVID is being approached in the US as an identity issue, like abortion or gun ownership. The language and discourse used precludes debate and the provision of new evidence, perspectives or proposals in this unprecedented and fast evolving issue with many remaining unknowns.

This research is, of course, not without limitations. It is based on a single week of analysis; a single social media platform (Twitter), whose users do not represent the wider US population; and only a very small sample of the huge number of posts made in relevant trending topics during the period under consideration. The findings of this research should not be an end-point but can, perhaps, form a basis for broader and more in-depth research going forward.

However, it is hoped that the findings and analysis herein can form some temporary guidance and indications for those who might hope that social media could help understand public reactions to restriction measures and the reasons underpinning these reactions, and help facilitate information sharing, debate and discussion.

The first major point here concerns how these issues are discussed, particularly by supporters. The idea of rights, freedoms and responsibilities as a complicated balance is being lost in this debate. Supporters would likely be more successful in engaging in debates if they discussed ideas of community rights and social responsibilities. Arguments of following state and official guidance are lost on those who see individual rather than community rights and COVID as a hoax by an establishment.

We should also pay more attention to the fact that the rights and responsibilities that underpin COVID restrictions are fundamentally based around a collective or socially orientated set of rights: freedom from disease, especially for the most vulnerable; freedom from unprecedented health consequences in ill-equipped workplaces; collective individual changes of action to support vulnerable members of society. These practical attempts to limit the spread of COVID in Western counties have largely been undertaken based on practices first trialled in East Asian societies, with stronger collective values. The local lockdown model also originated in mainland China, where there is a very strong existing state system of wide-spread monitoring and enforcement.

To the extent that we want to discuss these restrictions we need both to adopt and critically evaluate these, somewhat less familiar, although not foreign collective rights discourses. We should learn how these arguments and articulations are being deployed in these contexts, while also engaging in debates about how these arguments and articulations might need to be modified in systems that are based on a different cultural interpretation of rights and freedoms[7].

This brings us to the second major point: polarisation. These debates are precluded if approaches to COVID are polarised along existing political identity positions that are resistant to evidence and debate. Rather than debating about the people who hold these opinions (patriots or terrorists, selfish idiots or sheep) and the people who propose opinions and policies (democrats or republicans), we need to focus in this fast-moving and unprecedented situation on debating the deep underlying principles, while remaining cognizant and open to the new information that will inevitably emerge.

This data memo has intended to provide a starting point for these debates by approaching this issue through a political science framework of different kinds of rights, freedoms and responsibilities and the many ways that these can be balanced and structured in different perspectives, cultures and contexts. It is hoped that this debate can be continued by academics, media and individuals as part of everyday political conversation.

Many hope that the debates and difficult choices precipitated by COVID will lead to major changes that will result in better societies, economies and political systems. The theoretical debates about balances of rights and freedoms undertaken here, as well as the practical, structural and intersectional application of these, will be important in setting the foundation for any positive societal changes that might be precipitated by this difficult situation.

---

[1] We note that some have contested this distinction, for instance, MacCallum who argues that all freedom is negative (i.e freedom from constraints) (MacCallum, 1967).

However, a deep discussion of these debates is outside the bounds of this short data memo.

[2] Given the prevalence of political commentary on COVID policies and practices in the dataset, expected agreement for this code was 94.4%. With such a high expected agreement, percentage agreement and percentage agreement exceeding expected agreement is a better measure of intercoder reliability than measures such as Kappa, which are too stringent in cases of very high expected agreement.

[3] In previous research we have noted three reasons for unrelated trends collected in the dataset. Firstly, this method collects the most recent 100 tweets that would show up in a search for the trending topic hashtag or keyword on Twitter. However, Twitter often will show unrelated popular or trending content alongside a search, which would also be collected. Secondly, some keywords take in a variety of information. For instance, the trending keyword "michiganders" was coded as being about the Michigan protest based on a majority of the content in the trend but could have been used in the context of non-political, non-COVID discussions. Lastly, irrelevant tweets also are collected due to the practice of using trending hashtags and words unrelated to the content of the tweet to gain exposure. These results suggest that for more prominent trends the extent of focus on the topic is less than for less prominent trends. This might suggest the third explanation is most accurate, as fraudulent hashtag or keyword users would tend to focus on the most popular trends to spread their unrelated messages. However, this explanation is unsatisfactory as a reading of the selected tweets rarely shows this practice occurring overtly. For instance, the tweet "Being friends with me, You gotta understand I do not wanna talk EVERYDAY 😂" was collected when the term "capitol" was searched. There is no use of the word capitol or any synonyms in the tweet, nor in the usernames, profiles or descriptions of either the posting user or retweeting user. This finding suggests the possibility that searches for more popular trending queries might be more likely to include other popular but unrelated information than searches for less popular queries. Although it is impossible to know the inner workings of the Twitter search queries, this would not be illogical as someone searching for the keywords amazon, america or biden (the three largest trends across the period) would be more likely to be tolerant, or even appreciative, of irrelevant content in their search results than someone searching for less popular trends such as nypd, sonic and bollywood, each of which only occurred once in one of the 64 US locations during the period.

[4] In stating media positioning, we use the definition of AllSides, which provides an assessment of media bias based on multiple methods including blind surveys, editorial review, third party and independent analysis and community feedback on bias ratings.

[5] Although outside the bounds of this short data memo, the correlation of collectively orientated cultures and trust in the state across the world to the nature and success of COVID restriction policies would be fertile ground for future research.

[6] This is in contrast to mainland China were until more recently mask wearing has been largely to protect oneself from airborne pollutants.

[7] This is of course entirely precluded by the worryingly prevalent conspiracy theory that COVID was "invented" by China for political purposes.

**Acknowledgements**

I wish to thank Emma Briones, Rhian Hughes and Janet Tokitsu Tizon for their content analysis work on these data; the University of Leeds Strategic Research Investment Fund for helping pay for Emma Briones' and Rhian Hughes' work on the project; and the Laidlaw Scholarship program for funding and facilitating Janet Tokitsu Tizon's work on the project. I also wish to thank Google for providing Cloud Computing research credits that allowed data collection and processing to be undertaken for this research at a time when research funding has been severely curtailed due to COVID-related financial uncertainty.

**References**

Auxier, B. (2020, May 4). How Americans see digital privacy issues amid the COVID-19 outbreak. *PEW Research Centre*. https://www.pewresearch.org/fact-tank/2020/05/04/how-americans-see-digital-privacy-issues-amid-the-covid-19-outbreak/

BBC. (2020a, January 23). *China coronavirus: Lockdown measures rise across Hubei province*. https://www.bbc.co.uk/news/world-asia-china-51217455

BBC. (2020b, March 15). *Coronavirus: Spain and France announce sweeping restrictions*. https://www.bbc.co.uk/news/world-europe-51892477

Beckett, L. (2020, April 30). Armed protesters demonstrate against Covid-19 lockdown at Michigan capitol. *The Guardian*. https://www.theguardian.com/us-news/2020/apr/30/michigan-protests-coronavirus-lockdown-armed-capitol

Berlin, I. (1969). Two concepts of liberty. In *Four Essays On Liberty* (pp. 18–172). Oxford University Press.

Bolsover, G. (2017). *Technology and political speech: Commercialisation, authoritarianism and the supposed death of the Internet's democratic potential* [University of Oxford]. https://ora.ox.ac.uk/objects/uuid:f63cffba-a186-4a6c-af9c-dbc9ac6d35fb/download_file?file_format=pdf&safe_filename=__Bolsover_PhD_6_for%2Bprinting%2Band%2Bbinding.pdf&type_of_work=Thesis

Bolsover, G. (2018). Slacktivist USA and Authoritarian China? Comparing Two Political Public Spheres With a Random Sample of Social Media Users: Microblog Users in the United States and China. *Policy & Internet*, *10*(4). https://doi.org/10.1002/poi3.186

Bolsover, G., & Tokitsu Tizon, J. (2020). *Social Media and Health Misinformation during the US COVID Crisis*. Centre for Democratic Engagement, University of Leeds. https://cde.leeds.ac.uk/wp-content/uploads/sites/93/2020/08/20200722-Social-Media-and-Health-Misinformation-during-the-US-COVID-Crisis.pdf

Broom, D. (2020). *People in France can't leave home without this permit during the COVID-19 outbreak*. World Economic Forum. https://www.weforum.org/agenda/2020/03/people-in-france-can-t-leave-home-without-this-permit-during-the-covid-19-outbreak/

CDC, C. for D. C. and P. (2020). *Coronavirus Disease 2019: Cases in the U.S.* https://www.cdc.gov/coronavirus/2019-ncov/cases-updates/cases-in-us.html

EFF. (2020). COVID-19 and Digital Rights. *Electronic Frontier Foundation*. https://www.eff.org/issues/covid-19

Feng, E., & Chen, A. (2020, February 21). Restrictions And Rewards: How China Is Locking Down Half A Billion Citizens. *NPR*. https://text.npr.org/s.php?sId=806958341

Hedgecoe, G. (2020, April 26). Coronavirus: Spain's children run free from lockdown—But not all. *BBC*. https://www.bbc.co.uk/news/world-europe-52409407

Hofstede, G., Hofstede, G. J., & Minkov, M. (2010). *Cultures and organizations: Software of the mind; intercultural cooperation and its importance for survival*. McGraw-Hill.

Jones, O. (2020, May 14). The real message behind 'stay alert': It'll be your fault if coronavirus spreads. *The Guardian*. https://www.theguardian.com/commentisfree/2020/may/14/stay-alert-coronavirus-blame

Lombard, M., Snyder-Duch, J., & Bracken, C. C. (2002). Content analysis in mass communication: Assessment and reporting of intercoder reliability. *Human Communication Research*, *28*(4), 587–604.

MacCallum, G. (1967). Negative and Positive Liberty. *The Philosophical Review*, *76*(3), 312–334.

Orange, R. (2020, March 28). As the rest of Europe lives under lockdown, Sweden keeps calm and carries on. *The Guardian*. https://www.theguardian.com/world/2020/mar/28/as-the-rest-of-europe-lives-under-lockdown-sweden-keeps-calm-and-carries-on

Perry, E. J. (2008). Chinese Conceptions of: From Mencius to Mao—and Now. *Perspectives on Politics*, *6*(01), 37–50. https://doi.org/10.1017/S1537592708080055

Portes, J. (2020, March 25). Don't believe the myth that we must sacrifice lives to save the economy. *The Guardian*. https://www.theguardian.com/commentisfree/2020/mar/25/there-is-no-trade-off-between-the-economy-and-health

Rambaran, V. (2020, April 22). Most Americans support extending coronavirus stay-at-home orders, poll finds. *Fox News*. https://www.foxnews.com/us/americans-support-extending-coronavirus-stay-at-home-orders

Selin, J. (2020, April 17). Trump versus the states: What federalism means for the coronavirus response. *The Conversation*. https://theconversation.com/trump-versus-the-states-what-federalism-means-for-the-coronavirus-response-136361

Zhou, D. X. (2010). Measuring the meaning of political concepts in Chinese online deliberation. *Avhandling, Hämtad*, *3*.